\documentclass[useAMS,usenatbib,letter]{mn2e}

\usepackage{graphicx}
\usepackage{multirow}
\usepackage{natbib}
\newcommand{\aap}{A\&A}
\newcommand{\apj}{ApJ}
\newcommand{\apjl}{ApJL}
\newcommand{\mnras}{MNRAS}
\newcommand{\ba}{\begin{eqnarray}}
\newcommand{\ea}{\end{eqnarray}}
\newcommand{\be}{\begin{equation}}
\newcommand{\ee}{\end{equation}}

\title[How strong are the Rossby vortices?]{How strong are the Rossby vortices?}

\author[H. Meheut, R.V.E. Lovelace, D. Lai]{H. Meheut$^{1,2}$\thanks{E-mail:
heloise.meheut@cea.fr}, {R.V.E. Lovelace$^{3}$} and D. Lai$^{3}$\\
$^{1}$Physikalisches Institut \& Center for Space and Habitability, Universit\"at Bern, 3012 Bern, Switzerland\\
$^{2}$CEA, Irfu, SAp, Centre de Saclay, F-91191 Gif-sur-Yvette, France\\
$^{3}$Department of Astronomy, Cornell University, Ithaca, NY 14853, USA}
\begin{document}

%\date{Accepted 1988 December 15. Received 1988 December 14; in original form 1988 October 11}

\pagerange{\pageref{firstpage}--\pageref{lastpage}} \pubyear{2012}

\maketitle

\label{firstpage}

\begin{abstract}
The Rossby wave instability, associated with density bumps in 
differentially rotating discs, may arise in several different
astrophysical contexts, such as galactic or protoplanetary discs. 
While the linear phase of the instability has been well studied, the
nonlinear evolution and especially the saturation phase remain poorly
understood.
In this paper, we test the non-linear saturation mechanism analogous to that derived
for wave-particle interaction in plasma physics. To this end we perform
global numerical simulations of the evolution of the instability in a
two-dimensional disc.
We confirm the physical mechanism for the instability 
saturation and show that the maximum amplitude of vorticity can be
estimated as twice the linear growth rate of the instability. We provide
an empirical fitting formula for this growth rate for various parameters of the density bump.
 We also investigate the effects of the azimuthal mode number of the instability
and the energy leakage in the spiral density waves. Finally, we show 
that our results can be extrapolated to 3D discs.
\end{abstract}


\section{Introduction}

Understanding the evolution of accretion discs is important for several astrophysical applications. The Rossby wave instability 
(RWI) \citep{LOV99} and Rossby vortices it form have been
discussed in a large variety of disc systems, from large-scale 
galactic discs \citep{LOV78,SEL91} and the Galactic centre
\citep{TAM06}, to discs in microquasars \citep{LTR09}
and protoplanetary discs \citep{VAR06,MKC12}.
The latter are of particular importance since Rossby vortices may 
accelerate planetesimal formation.  
The RWI operates around the density bumps or vortensity bumps
(see Section 2) in the disc.
In the linear theory, the growth rate of the instability
can be computed as an eigenvalue problem. This was done by \citet{LI00} 
for 2D discs, by \citet{U10} in the shallow water approximation, and by 
\citet{MYL12} and \citet{L12} for 3D vertically stratified discs.
After the linear phase, the instability saturates and the linear theory
breaks down. The saturation amplitude of the instability is of
importance in the several applications.
For example, in protoplanetary discs, the vortices formed by the RWI can concentrate 
the solid particles and hence accelerate the formation of planetesimals. 
The amplitude of the vortices is a key parameter that determines 
the amount of solids that can be concentrated and the size of the resulting planetesimals
\citep{MMV12}. The Rossby vortices can also influence 
planet migration in the disc in the latter stage of planet formation
\citep{KLL03}, and the amplitude of migration is determined by 
the strength of the vortices.
In the RWI model of Quasi-Periodic Oscillations (QPO) in microquasars,
the saturation amplitude of the RWI is of importance
for the understanding of the observed QPO amplitude.

The saturation mechanism of the RWI was already discussed in
\citet{LI01}, but no criterion for saturation was given. In this
paper, we interpret the growth of the instability in term of
particle-wave interaction and use classical plasma physics results in this context,
which give the non-linear saturation condition as proposed by
\citet{LTR09}.  In section 2, we present the physical
interpretation of the growth of the instability and the model for the
saturation mechanism. The numerical simulations are presented and
compared with the model in section \ref{Sec:simu}.
We then summarise and conclude in the last section.

\section[]{The RWI}

\subsection{Linear growth}

The RWI can be seen as an equivalent of the Kelvin-Helmholtz instability 
in differentially rotating discs, and has a similar instability criterion: an extremum 
in the quantity $\mathcal L$ related to the vorticity of the equilibrium flow. In a 
non-magnetised thin disc, this quantity can be written as
\begin{equation}
\mathcal L= \frac{\Sigma \Omega}{\kappa^2} (p\Sigma^{-\gamma})^{2/\gamma} =	
\frac{\Sigma}{2(\vec\nabla\times\vec v)_z}(p\Sigma^{-\gamma})^{2/\gamma} \,,
\end{equation}
where $\Sigma$ is the surface density, $\vec v$ the velocity of the fluid, 
$\gamma$ the adiabatic index, $\Omega$ the rotation frequency, and 
$\kappa^2 = 4\Omega^2 + 2r\Omega\Omega '$ the squared epicyclic frequency 
(so that $\kappa^2/2\Omega$ is the vorticity). Here the prime denotes a radial 
derivative. 
For barotropic discs considered in this paper, 
the quantity ${\mathcal L}$ reduces to
\be
\mathcal L=\frac{\Sigma \Omega}{\kappa^2}=\frac{\Sigma}{2(\vec\nabla\times\vec v)_z},
\ee
corresponding to $1/2$ of the inverse 
vortensity\footnote{Vortensity is defined as the 
ratio of vertical vorticity to surface 
density:$\frac{(\vec\nabla\times\vec v)_z}{\Sigma}$}.

%============Figure:schema_RWI
\begin{figure}
   \centering
   \includegraphics{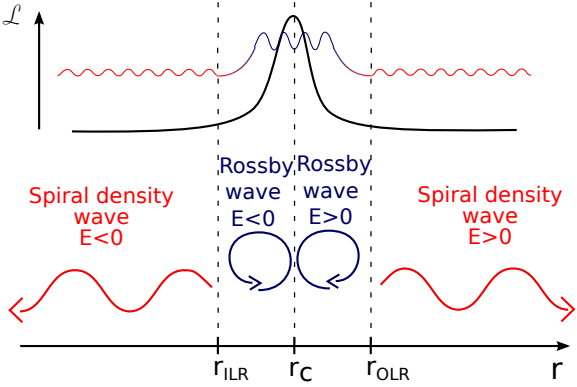}
      \caption{Schematic view of the RWI with the two propagating
        regions for the Rossby waves and density waves, and in between
        the evanescent regions. At the corotation radius $r_c$ the wave
        pattern frequency matches the orbital frequency
        (Eq. \ref{eq:corotation}).  The inner and outer Lindblad
        radii, $r_{\rm ILR}$ and $r_{\rm OLR}$, are defined by
        Eq.~(\ref{eq:lindblad})}.
         \label{Fig:schema_RWI}
\end{figure}
%===============================

In a disc with such an extremum, two standing Rossby waves can
coexist, one on each side of the extremum.
We recall here that Rossby
waves are vorticity waves propagating on the gradient of vortensity, and
their dispersion relation is given by
\be
\tilde\omega=-\frac{2\kappa^2c_s^2}{c_s^2(k_r^2+m^2/r^2)+\kappa^2}\frac{m\mathcal L'}{r\Sigma}
\ee
where $\omega$ is the wave frequency, $\tilde \omega\equiv\omega-m\Omega$
is the wave frequency in the rotating frame, $m$ and $k_r$ its azimuthal 
and radial wave numbers, and $c_s$ is the sound speed. 
The wave corotation radius $r_c$ is defined by:
\be
\tilde\omega(r_c)\equiv\omega-m\Omega(r_c)=0.\label{eq:corotation}
\ee
Thus, with the negative gradient of the disc rotation velocity, we have
$\tilde\omega_{r>r_c}<0$ and $\tilde\omega_{r<r_c}>0$. From the
dispersion relation, one can see that Rossby waves propagate only
in the region where $\mathcal L'>0$ if $r<r_c$ and where $\mathcal
L'<0$ if $r>r_c$.

We consider here the case of a maximum of $\mathcal L$ as plotted in
Fig.~\ref{Fig:schema_RWI}.
The wave located in the positive gradient of $\mathcal L$ ($r<r_c$) has 
the pattern frequency smaller than the disc rotation frequency ($\omega/m<\Omega$) 
and carries negative energy (i.e., increasing wave amplitude tends to
to remove energy from the fluid). On the other hand, the wave in the
negative gradient of $\mathcal L$ has $\omega/m>\Omega$ and carries positive energy.
The growth of the RWI is due to the interaction between
the two Rossby waves with respectively positive and negative energies. 
The total energy is conserved while the wave amplitude increases. 
The mechanism for the growth of the instability is then
localised in the corotation region and does not depend on the boundary
conditions.

As explained by \citet{TAG01}, in accretion discs the differential
rotation couples Rossby waves to density waves. These density waves
can only propagates outside the two (inner and outer) Lindblad
resonances radii $r_{\rm LR}$ defined as
\be
\omega-m\Omega(r_{\rm LR})=\pm \kappa(r_{\rm LR}).\label{eq:lindblad}
\ee
Between the two propagation zones the wave is evanescent and
tunnelling is possible as plotted in Fig.~\ref{Fig:schema_RWI}. As
detailed in \citet{TSA08} and \citet{LAI09}, these propagation regions
clearly appear when expressing the linearized perturbation equations
in the form of Schr\"odinger equation, with the effective potential given by
  \be
  V_{\rm eff}=\frac{2m\Omega}{r\tilde \omega}\frac{d}{d_r}\Big(\ln\frac{\Omega\Sigma}{\kappa^2-\tilde\omega^2} \Big)+\frac{m^2}{r^2}+\frac{\kappa^2-\tilde\omega^2}{c_s^2}. 
  \ee
The waves are propagating in the region where $V_{\rm eff}(r)$ is
negative and evanescent where $V_{\rm eff}(r)$ is positive. This effective
potential is plotted in Fig.~\ref{Fig:potentiel} for the specific
configuration considered below.

\subsection{Non-linear saturation}\label{sec:sat}

To estimate the amplitude at which the linear calculation stops to be
valid, we use a standard plasma criterion which gives, for instance,
the breakdown of the Landau damping theory due to particle trapping
\citep{ONE65,KT73}. We consider a fluid particle in a Rossby wave with the 
vorticity amplitude $\omega_v$ exponentially growing with time during the
linear stage,
\be 
|\omega_v|\propto \exp(\gamma t).
\ee

We here use the convention to express time in units of $\Omega_0^{-1}$.

The estimation of the rotation time of a fluid particule in a vortex is related to the vorticity. For instance for a circular vortex with constant vorticity, it can be estimated as
\be
\tau_T\sim 4\pi/|\omega_v|.
\ee

However in the case of the vortices formed by the RWI, the vortex doesn't have an analytical expression and the turnover timescale can not be computed analytically. From the simulations presented in the next section, we estimated
\be
\tau_T\sim 2/|\omega_v|.
\ee

This circular motion around the vortex centre is a major perturbation
to the circular orbit of the fluid particle around the disc centre and
it modifies the radial structure of the disc in the region of the
vortices on a time scale $\tau_T$.  The linear calculation of the
growth of the Rossby wave instability is determined by this radial
structure of the disc and especially by the radial structure of
$\mathcal L$. Therefore, this linear approach is only valid 
when the growth timescale ($1/\gamma$)
is smaller that $\tau_T$.  This occurs when
\be
\frac{|\omega_v|}{2}\sim\gamma.
\label{eq:sat}
\ee
In other words, the exponential growth of the wave amplitude
necessarily ends when the fluid particle motion in the
vicinity of resonant radius $r_c$ differs appreciably from the
strictly azimuthal motion in the equilibrium.  
Note that the saturation criterion (\ref{eq:sat}) is only order-of-magnitude, and similar criterion was proposed by \citet{LTR09} in a different context. Below we perform non-linear numerical simulations to calibrate this criterion.

\section{Non-linear simulations}\label{Sec:simu}

\subsection{Methods}

We have performed two dimensional (2D) non-linear simulations of a
differentially rotating disc with a constant surface density except
for a Gaussian bump. This over-density gives an extremum of $\mathcal
L$ in which the instability can grow.

We work in cylindrical coordinates $(r, \varphi)$ with the 2D Euler equations
\begin{eqnarray}
\partial_t\Sigma+\vec \nabla\cdot (\rho\vec v)=0 \,,\\
\partial_t(\Sigma\vec v)+\vec\nabla\cdot(\vec v\Sigma\vec v)+\vec\nabla p=-\Sigma\vec\nabla \Phi_G \,,
\end{eqnarray}
where $\Sigma$ is the surface density of the fluid, $\vec v$ its
velocity, and $p$ its pressure.  
$\Phi_G = -GM_*/r$ is the gravitational potential of the central object with $G$ the
gravitational constant and $M_*$ the mass of the star.  We consider a
\emph{globally} isothermal flow, i.e.\ with a radially constant sound
speed, and a linear relation between pressure and density $p =
c_s^2\Sigma$.

The initial surface density, normalized to $\Sigma_0$ is given by
\begin{equation}
\Sigma / \Sigma_0=1+\chi \exp\Big(-\frac{(r-r_0)^2}{2\sigma^2}\Big).
\label{Eq:dens}
\end{equation}
Our canonical parameters are $r_0=1.$, the amplitude of the bump $\chi$ is chosen in the range $[0.15,0.3]$, and its width is given by $\sigma/r_0=0.05$. Although we will also present results for different values of $\sigma/r_0$. The disc scale height is fixed to $h=c_s/\Omega_0=0.1r_0$,
$\Omega_0$ being the Keplerian orbital frequency at $r_0$. 
The initial azimuthal velocity is chosen to achieve radial equilibrium. A low 
amplitude perturbation is added to this equilibrium:
\begin{equation}
v_r=\epsilon \sin(m\phi)\exp\Big(-\frac{(r-r_0)^2}{2\sigma^2}\Big)
\end{equation}
with $\epsilon=10^{-4}$ and the azimuthal mode number $m\in[2,5]$. 
We have also performed simulations with initial random perturbations of the 
same amplitude.

We use the Message Passing Interface-Adaptive Mesh Refinement
Versatile Advection Code (MPI-AMRVAC) \citep{KEP11}, with a
uniform grid. The numerical scheme is the Total Variation Diminishing
Lax-Friedrich scheme (see \citealt{TOT96}) with a third order accurate
Koren limiter \citep{KOR93} on the primitive variables. We use a
uniform cylindrical grid with $r/r_0\in[0.2, 2.4]$, and the full
azimuthal direction $\varphi\in[0, 2\pi]$. Numerical convergence has
been tested and two different resolutions have been used: $(1024,128)$
for most of the simulations ($m=1$ to $4$) but a resolution of
$(1024,256)$ was needed for the highest azimuthal modes number $m=5$ and the
simulations with white noise perturbations. 
There is a null radial velocity at the inner and outer boundary 
of the grid.

%============Figure:potentiel
\begin{figure}
   \centering
	\includegraphics[width=9cm,clip=true]{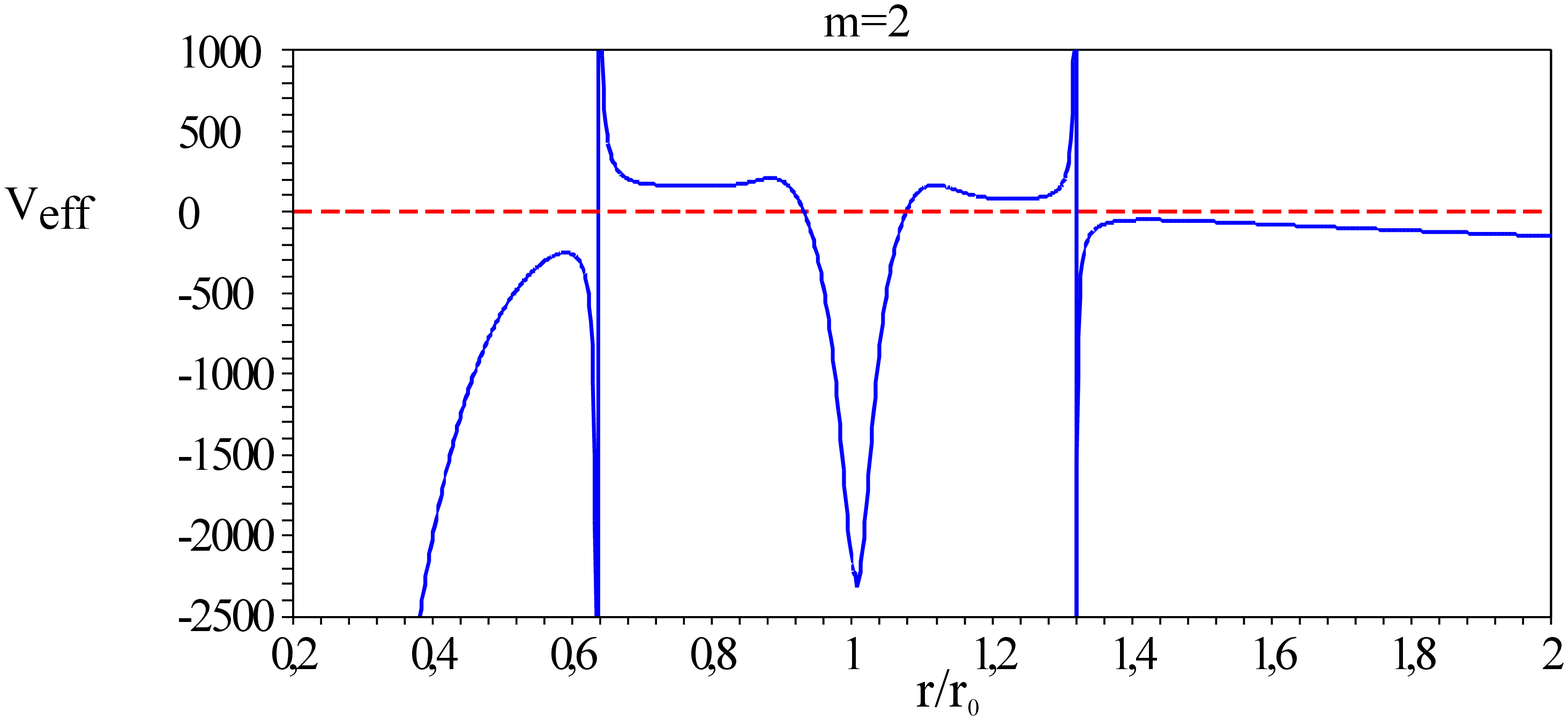}
  \vspace{-0.1cm}
   \includegraphics[width=9cm,clip=true]{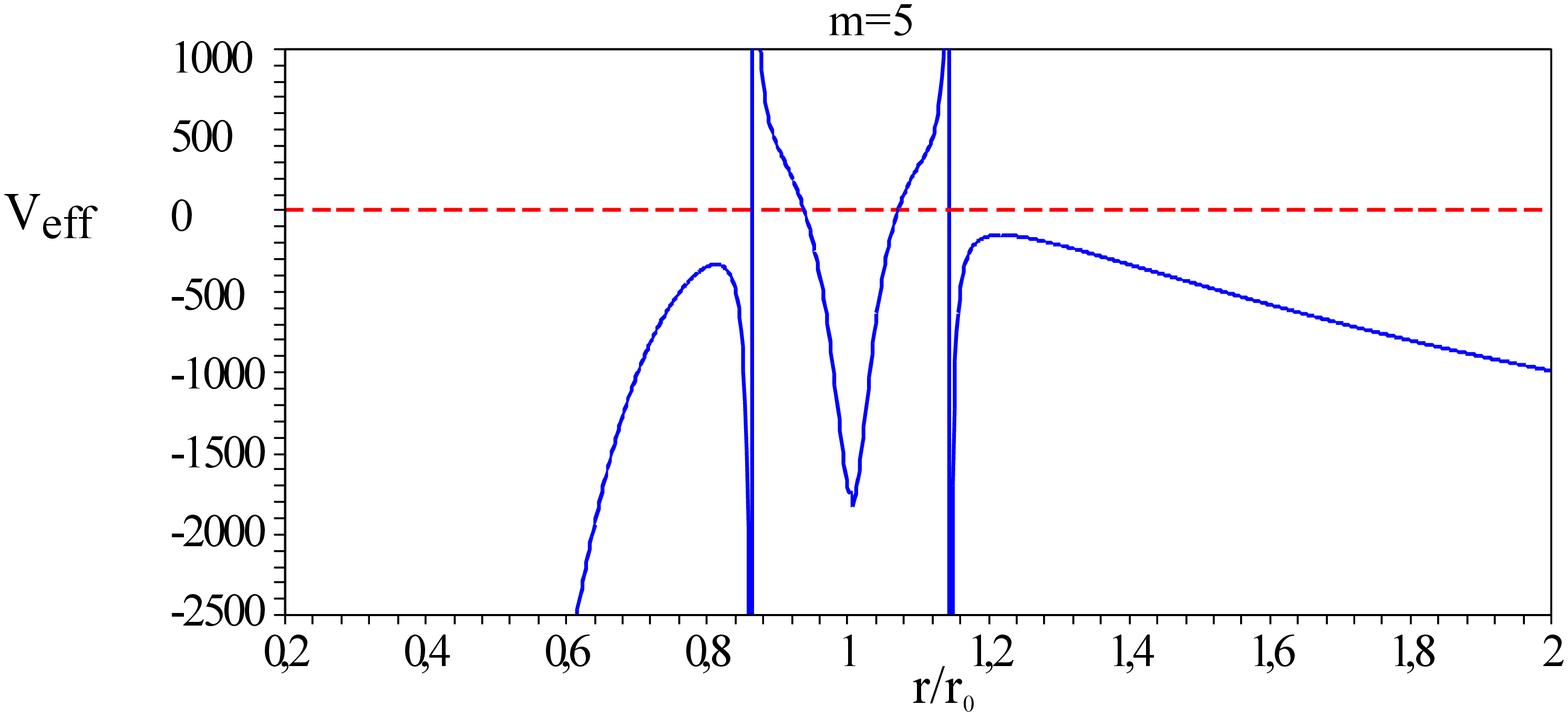}
      \caption{Effective potential for a density profile given by
        Eq.~\ref{Eq:dens} with $\chi=0.2$ for $m=2$ (upper panel) and
        $m=5$ (lower panel). Waves can propagate only in the regions
        where $V_{\rm eff}(r)<0$.}
         \label{Fig:potentiel}
\end{figure}
%===============================

This numerical configuration is very similar to the one of
\citet{MYL12} which has proven to accurately describe the instability,
with a very good agreement with the analytical approach in the linear
phase. For all these simulations, the square of the epicyclic
frequency, $\kappa^2$, is positive at all radii and the disc is stable 
against axisymmetric perturbations. 

\subsection{Results}

%============Figure:growths
\begin{figure*}
\includegraphics[width=18cm]{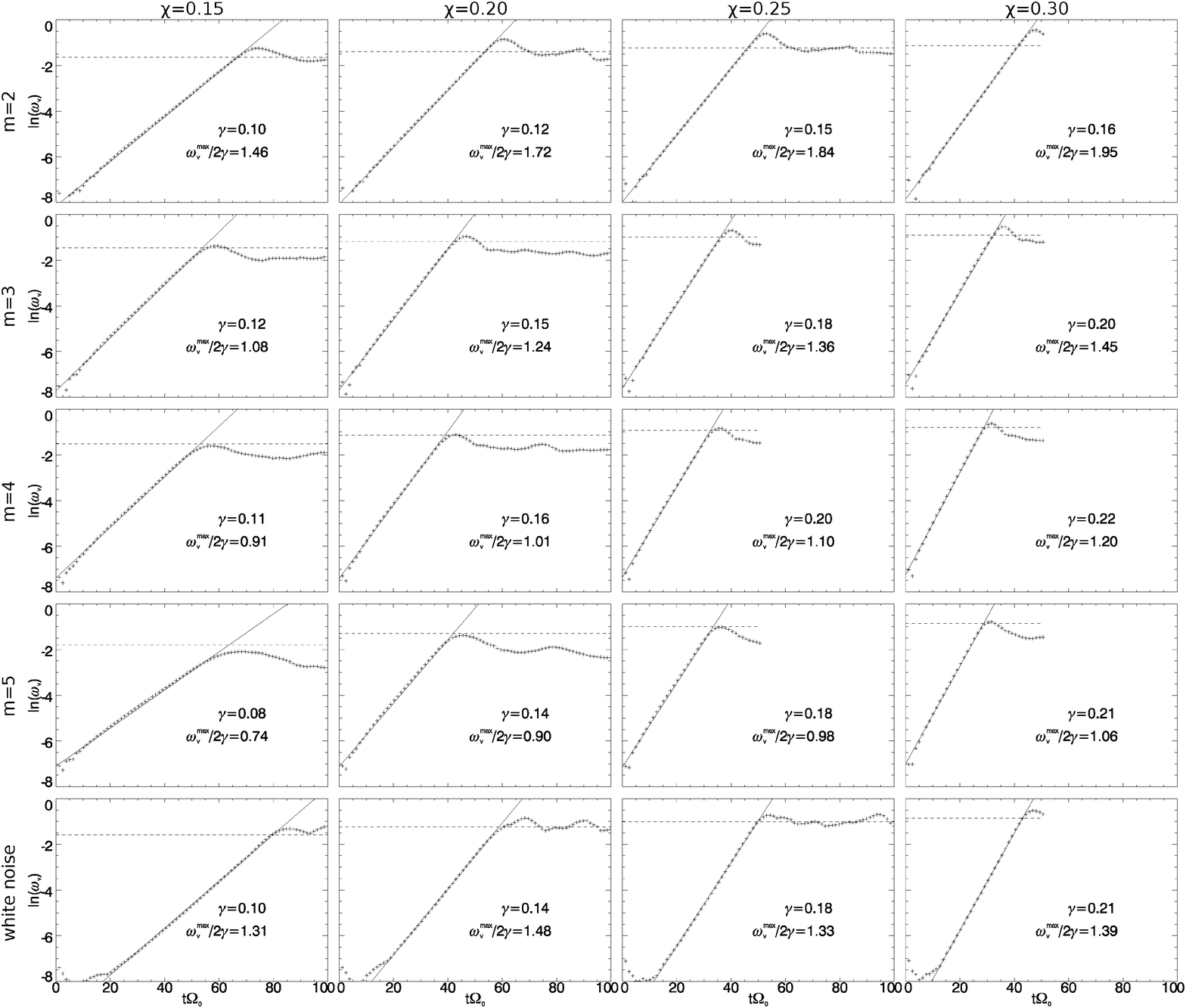}
      \caption{Amplitude of the Rossby waves (vorticity) on
logarithmic scale as a function of time (in units of $\Omega_0^{-1}$). A fit of
the exponential growth (solid line) gives the growth rate
$\gamma$. The dashed line corresponds to the saturation amplitude
estimated by the model. 
}
         \label{Figgrowths}
\end{figure*}
%===============================

For each of the $20$ simulations, the linear growth rate of the
instability is obtained by fitting the amplitude of the
vorticity in the Rossby waves region as a function of time (see
Fig.~\ref{Figgrowths}). We have previously shown that this method
gives growth rate very similar to the one obtained by solving the
linear equations \citep{MYL12}.  The amplitude of the waves initially
grows exponentially until saturation. As explained in section
\ref{sec:sat}, we estimate saturation to arise when the waves reach a
vorticity of the order of $2\gamma$. This amplitude is plotted as a
dashed line in Fig. \ref{Figgrowths}.

For each simulation, the ratio of the maximum of vorticity to the expected value, 
$\omega_v^{max}/2\gamma$, is given.% There is a very good
%agreement between the prediction of the model and 
The saturation
amplitude reached in the non-linear numerical simulations for $m=4$
and $m=5$. Similar to the non-linear evolution of the Landau damping,
we also obtain oscillations in the waves amplitude.

\begin{figure}
   \centering
   \begin{tabular}{llll}
   	\hspace{1cm}$m=2$&$m=3$&$m=4$&$m=5$
   	\vspace{-0.5cm}
   \\
   \hspace{-0.7cm}
   	\includegraphics[width=13.5cm,angle=90,trim=0.cm 10.4cm 0.5cm 11.5cm,clip=true]{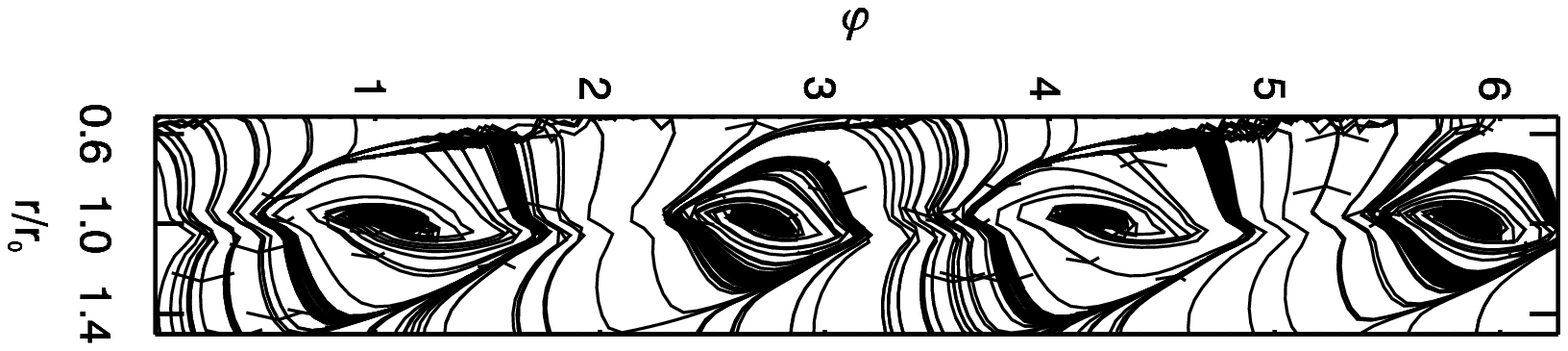}&
   	\hspace{-0.6cm}
   	\includegraphics[width=13.5cm,angle=90,trim=0cm 10.4cm 0.5cm 12.85cm,clip=true]{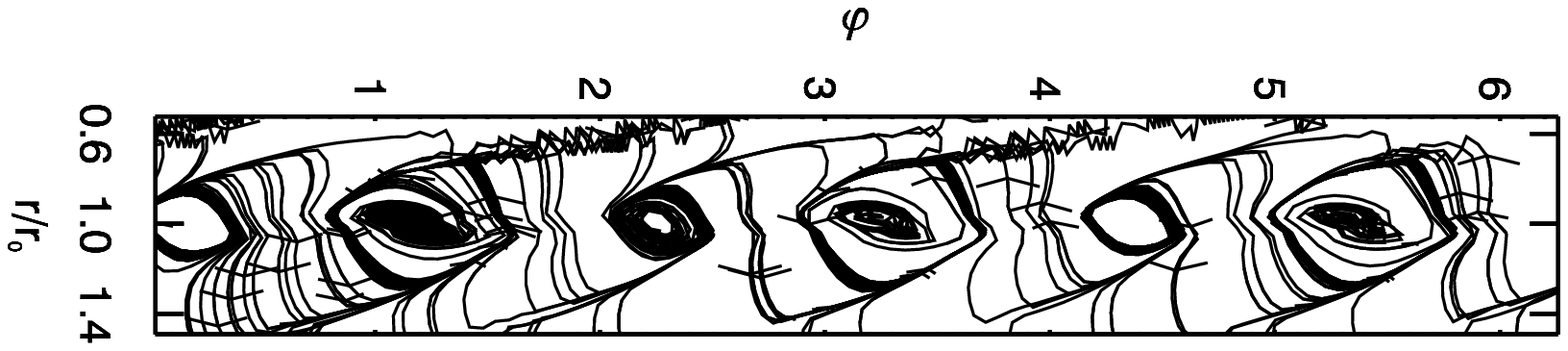}&
   	\hspace{-0.6cm}
   	\includegraphics[width=13.5cm,angle=90,trim=0cm 10.4cm 0.5cm 12.85cm,clip=true]{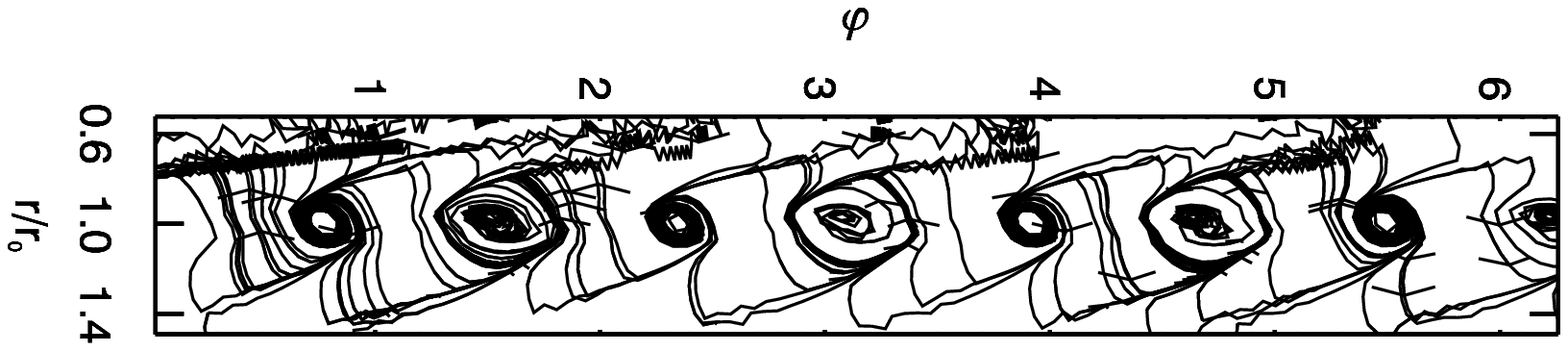}&
   	\hspace{-0.6cm}
   	\includegraphics[width=13.5cm,angle=90,trim=0cm 10.4cm 0.5cm 12.85cm,clip=true]{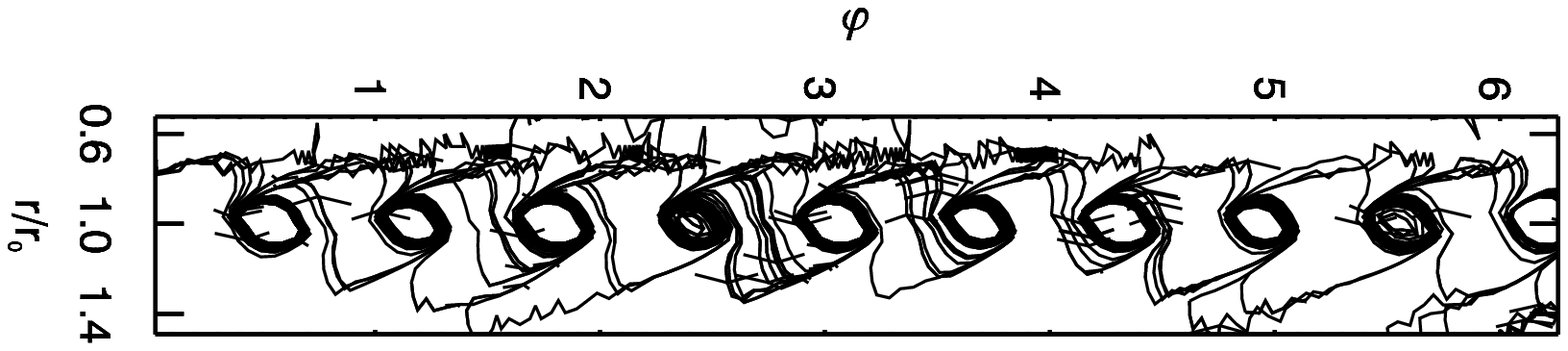}
   \end{tabular}
      \caption{Streamlines showing the shape of the vortices for the
        four azimuthal mode numbers considered here.
        For the lowest mode number the
        vortices are elongated, for the highest ones the vortices are
        circular.  }
         \label{Figstream}
\end{figure}

For low mode numbers, the maximum
vorticity is slightly below the estimation. This is due to the shape of the vortices. Indeed 
%we considered circular vortices to estimate the circulation time inside
%the vortex, but 
the vortices with low $m$ have elongated shapes as
can be seen in Fig.~\ref{Figstream} where vortices streamlines in the
$(r,\varphi)$ plane are plotted. These streamlines are computed with the
perturbed velocity during the linear stage of the instability. To allow for
a comparison between different $m$ cases, 
the time slices for these plots are chosen such that the perturbations have about the same amplitudes.
This shape was to be expected as the
width of the Rossby wave propagation region is fixed by the width of
the initial density bump and the length of the vortices is directly
related to the azimuthal mode number $m$. Assuming a doubled
circulation time for the elongated $m=2$ vortices gives the correct
saturation amplitude, as one can see in Fig.~\ref{Figgrowths} for the
$[m=2,\chi=0.30]$ simulation.

%============Figure:stream
%===============================

\begin{figure}
   \centering
   \begin{tabular}{ll}
   \vspace{-0.5cm}
   	\hspace{1.7cm}$m=2$&\hspace{.7cm}$m=5$
   \\
   	\includegraphics[width=13.5cm,angle=90,trim=0cm 9.cm 0.5cm 11.5cm,clip=true]{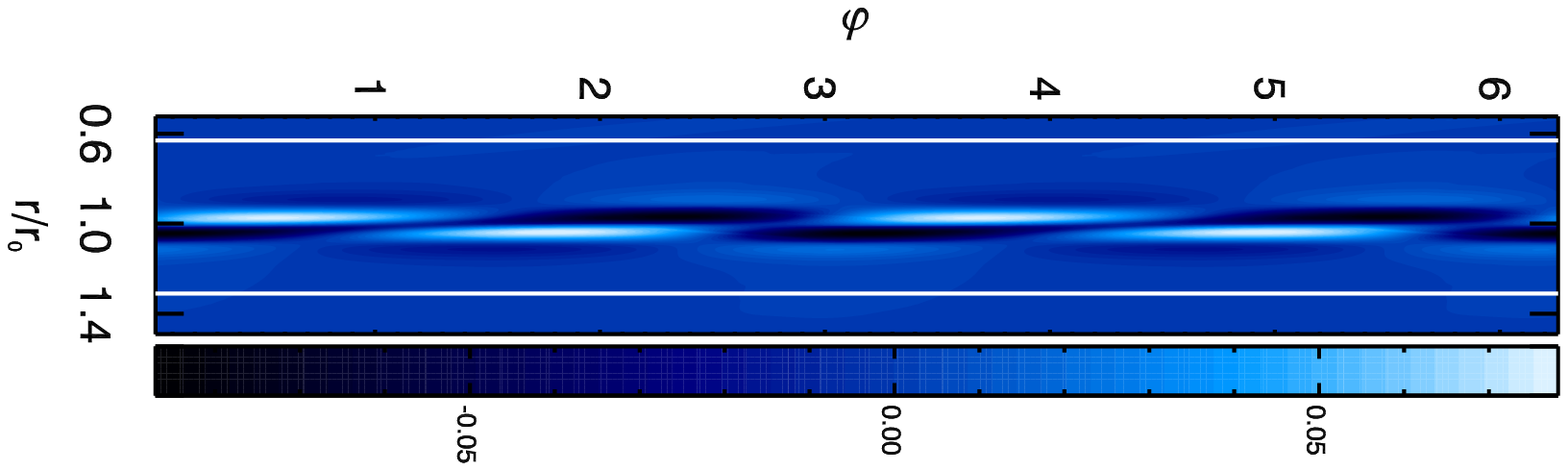}&
   	\includegraphics[width=13.5cm,angle=90,trim=0cm 9.cm 0.5cm 12.9cm,clip=true]{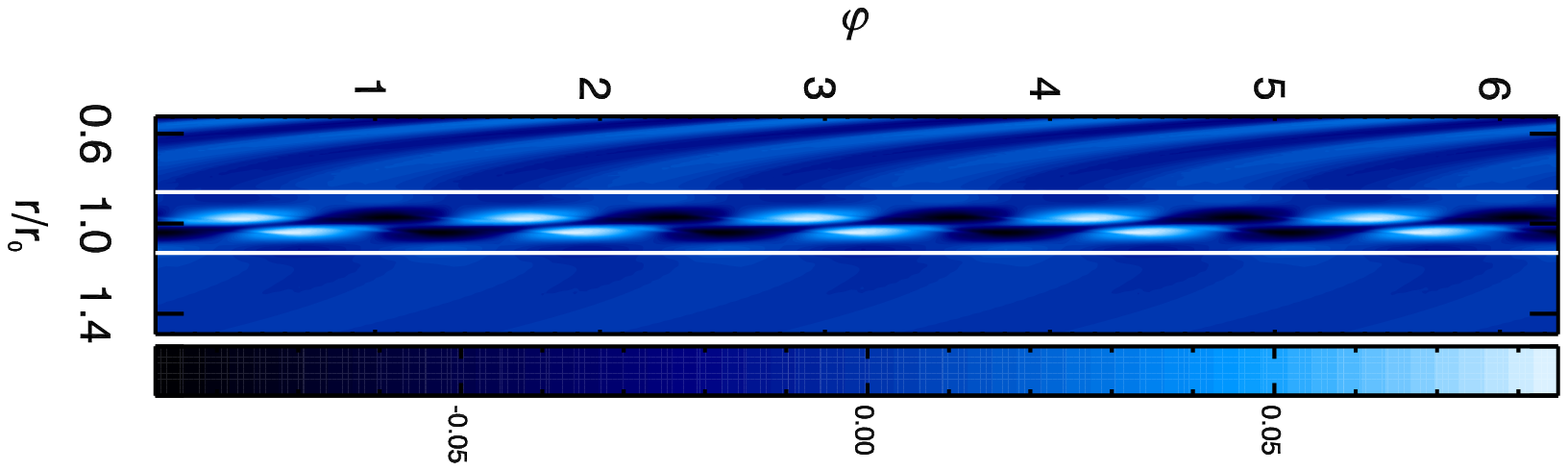}
   \end{tabular}
      \caption{Vorticity ($\Omega_0$) of the flow near corotation for
        $m2\chi25$ and $m5\chi25$. The white lines show the position
        of the Lindblad resonances. The width of the evanescent region
        between the vorticity wave and the Lindblad resonance is
        larger for lower $m$.  }
         \label{Figvortex}
\end{figure}

\begin{figure}
   \centering
	\includegraphics[width=6cm,angle=90,trim=1cm 1.0cm 2.cm 3.4cm,clip=true]{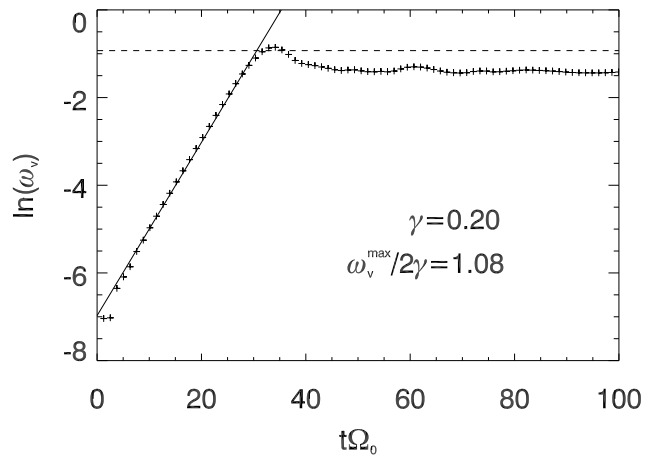}
      \caption{
      Amplitude of the Rossby waves (vorticity) on
	logarithmic scale as a function of time (in units of $\Omega_0^{-1}$) for the simulation
	with $\sigma/r_0=0.4$ and $m=5$. A fit of
	the exponential growth (solid line) gives the growth rate
	$\gamma$. The dashed line corresponds to the saturation amplitude
	estimated by the model. }
         \label{Fig:sigma}
\end{figure}

On the other hand, the vorticity maximum is overestimated for the
highest azimuthal mode number when the growth rate is low. See for
instance the $[m=5,\chi=0.15]$ simulation. We have checked that this is not
related to numerical dissipation by doubling the resolution and
obtaining no modification of $\omega_v^{max}$. Indeed the energy loss
responsible for the low saturation amplitude may be due to the density
waves propagating outside the Lindblad resonances that were not
considered in the local mechanism proposed for growth and saturation
in section \ref{sec:sat}. The amplitude of the density waves and the
energy loss is the highest when the Lindblad resonances are close to
the corotation radius and the width of the evanescent wave region is
small. This regions correspond to a positive effective potential
$V_{\rm eff}$ in Fig.~\ref{Fig:potentiel}. Since the Lindblad resonances
are closer to corotation for higher azimuthal mode number, the
evanescent region is narrower.
Energy transmission through this region is then carried away by density waves. 
These density waves can then become non-linear when they reach high amplitudes. This
may explain the lower amplitude at saturation. Moreover these density
waves are also responsible for the angular momentum transfer through
the disc and as a result for the evolution of the radial structure of
the disc from which the linear growth is computed. The distance
between the two propagation region appears clearly on
Fig.~\ref{Figvortex} where both the vorticity waves in the region of
corotation and the position of the Lindblad resonances are
visible. One can also identify the spiral density waves propagating
inwards and outwards from the Lindblad resonances in the $m=5$ plot.
To further check the importance of the Lindblad resonances, we have performed numerical simulations with a shallower density bump ($\sigma/r_0=0.4$) but with the same azimuthal mode number $m=5$. 
The density bump being shallower,
the Linblad resonances are situated further away from the vortices and less energy
is lost in the spirals waves. 
Fig.~\ref{Fig:sigma} shows that in this case the saturation amplitude is correctly estimated.

%============Figure:vortex
%===============================

%============Figure:sigma
%===============================

\begin{figure}
   \centering
   \includegraphics[width=6cm,angle=90,trim=1cm 1.0cm 2.cm 3.4cm,clip=true]{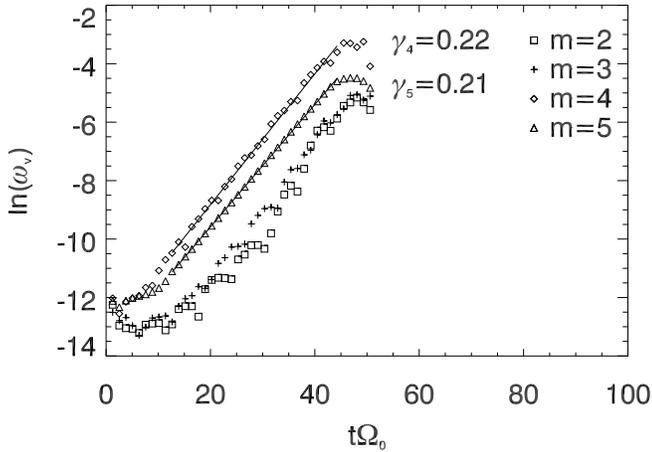}
   %height=8cm,angle=90,trim=1.cm 1.0cm 2.cm 2.4cm,clip=true]
      \caption{
      Amplitudes of the Rossby waves with different $m$'s as a function of time. The simulation starts out with a random white noise, with the parameters $\chi=0.3$, $\sigma/r_0=0.05$, $c_s/(r_0\Omega_0)=0.1$. The different $m$ component is obtained by a Fourier analysis in the azimuthal direction. The grow rates of the dominant modes are also given.}
         \label{Fig:growthmodes}
\end{figure}

The amplitude of saturation in the simulations with an initial white noise perturbation is more complex. Multiple modes
grow simultaneously in the linear phase with different growth rates, as can be seen on Fig.\ref{Fig:growthmodes}. However we still
 obtain an exponential growth of the total perturbation as the growth rates of the
 different modes are similar. The growth rate of the total perturbation ($0.21\Omega_0$) is slightly smaller than the highest growth rate of the different modes ($0.22\Omega_0$ for $m=4$) due to this coexistence of multiple modes with similar growth rate. These different modes then interact in the non-linear phase, but one can see that eq. (\ref{eq:sat}) still gives a good approximation of the saturation amplitude of the instability in this more realistic case.

%============Figure:growthmodes
%===============================

\subsection{Fitting formula}

In order to give an estimation of the growth rate of the instability based on the 
parameters of the density bump, we performed simulations with initial
random perturbations and with $\chi\in[0.1,0.3]$ and $\sigma/h\in[0.45,0.6]$. Fitting separately the dependence of the growth rate on $\chi$ and $\sigma/h$, and then combining, we find that $\gamma$ is approximately given by
\be
\frac{\gamma}{\Omega_0}\simeq\frac{k_1}{(\sigma/h)^{4/3}}\Big(\frac{\chi}{\sigma/h}-k_2\Big)^{2/3},
\label{eq:fit}
\ee
with $k_1=0.142$ and $k_2=0.138$. Here $\Omega_0$ the Keplerian frequency at the location of the density bump, $\sigma/h$ and $\chi$ specify the width and amplitude of the bump (see eq.~\ref{Eq:dens}), and $h=c_s/\Omega_0$ is the disc scale height.
Note that this fitting formula should be applied only in the range of parameters specified above and require that the disc be stable against axisymmetric perturbations (\i.e., the Rayleigh criterium $\kappa^2>0$, is satisfied). Eq.~\ref{eq:fit} also corresponds to the growth rate of the fastest growing mode ($m\sim 4$) in the linear theory.
The amplitude of maximum vorticity at saturation is given by
\be
|\omega_v|\sim\frac{2\Omega_0k_1}{(\sigma/h)^{4/3}}\Big(\frac{\chi}{\sigma/h}-k_2\Big)^{2/3}.
\ee

\subsection{3D simulation}

\begin{figure}
   \centering
   \includegraphics[width=6cm,angle=90,trim=1cm 1.5cm 2.cm 3.4cm,clip=true]{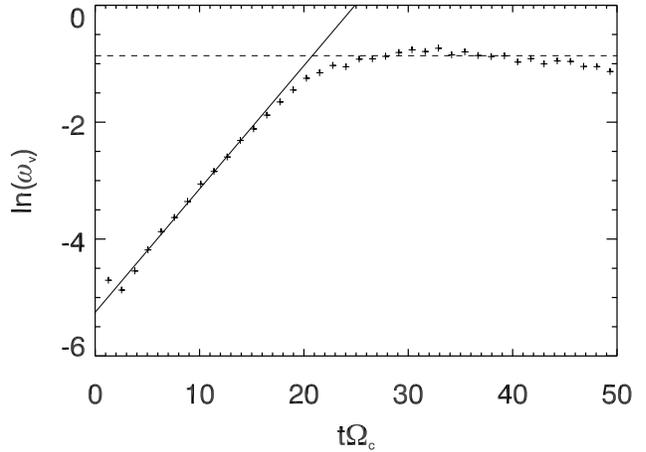}
      \caption{Amplitude of the Rossby waves (mid-plane vorticity) 
        on logarithmic scale as a function of time for the 3D
        simulation. A fit of the exponential growth gives the growth
        rate (solid line).}
         \label{Figgrowth3D}
\end{figure}

In the above we have considered 2D discs, neglecting their vertical structure. Recent study of the RWI in 3D vertically stratified discs has revealed unexpected vertical
displacements inside the vortices \citep{MEH10} that could be relevant
 for the saturation mechanism. However \citet{MYL12} and
\citet{L12} have shown that the growth rate is only slightly modified
(decreased) in a fully stratified disc. Moreover the vertical velocity
in the vortices is of low amplitude compared to the radial and
azimuthal perturbed velocity. As a result, we do not expect the
circulation time to be significantly modified by this vertical
circulation. To confirm the relevance of our model for 3D stratified
discs, we perform a numerical simulation in such a configuration. We
choose $m=4$ and $\chi=0.25$, and the volume density is chosen to have
a surface density similar to the $2D$ case (see \citealt{MYL12}). The
resulting growth and saturation of the instability is plotted in
Fig.~\ref{Figgrowth3D}. The predicted maximum amplitude of the
instability is correct, meaning that the model can be extended to 3D
discs. This can then be also applied to the saturation phase obtained 
in \citet{MKC12} where the long-term stability of the 3D vortices was
studied. After the saturation the vortices continue to evolve, they 
tend to merge as in an inverse cascade and eventually to decay.

%============Figure:growths
%===============================

\section{Summary}

We have tested numerically the non-linear saturation mechanism for the
Rossby wave instability proposed by \citet{LTR09}. This is based on
the classical results of particle-wave interaction well known for unstable
electron plasma waves. We have shown that non-linear saturation of the
RWI occurs when the trapping frequency of the particles in the Rossby
vortices equals the instability growth rate. We have estimated the
trapping frequency by the vorticity of the Rossby wave;
this is accurate for circular vortices but is only a lower limit for
elongated vortices. We also discussed the linear coupling between the
Rossby waves and the density waves. This has the effect of decreasing 
the amplitude of the vorticity wave. Finally, we gave an estimation
for the amplitude of the vortices at saturation based on the characteristics of the density bump.

Our paper focused on the non-linear saturation of the linear
instability in an initially steady disc. We did not considered the
mechanism for the formation of the pressure bump causing the
instability. In a disc where the bump is continuously sustained by
some accretion processes, the disc is not steady and a future step is to take
into account the evolution of the shape of the bump in the study of
the instability.

\section*{Acknowledgments}
We thanks the referee G. Lesur for his useful comments.
This work has been supported in part by NSF grants AST-1008245 and AST-1211061
and the Swiss National Science Foundation. RVEL was partly supported by NASA grant NNX11AF33G.

\label{lastpage}


\begin{thebibliography}{}

\bibitem[\protect\citeauthoryear{Keppens, Meliani, van Marle, Delmont, Vlasis
  \& van~der Holst}{Keppens et~al.}{2012}]{KEP11}
Keppens R.,  Meliani Z.,  van Marle A.,  Delmont P.,  Vlasis A.,    van~der
  Holst B.,  2012, Journal of Computational Physics, 231, 718

\bibitem[\protect\citeauthoryear{{Koller}, {Li} \& {Lin}}{{Koller}
  et~al.}{2003}]{KLL03}
{Koller} J.,  {Li} H.,    {Lin} D.~N.~C.,  2003, \apjl, 596, L91

\bibitem[\protect\citeauthoryear{{Koren}}{{Koren}}{1993}]{KOR93}
{Koren} B.,  1993, Notes on numerical fluid mechanics.
Vol.~45 of A robust upwind discretization method for advection, diffusion and
  source terms

\bibitem[\protect\citeauthoryear{{Krall} \& {Trivelpiece}}{{Krall} \&
  {Trivelpiece}}{1973}]{KT73}
{Krall} N.,  {Trivelpiece} A.,  {1973}, {Principles of plasma physics}.
{McGraw-Hill}

\bibitem[\protect\citeauthoryear{{Lai} \& {Tsang}}{{Lai} \&
  {Tsang}}{2009}]{LAI09}
{Lai} D.,  {Tsang} D.,  2009, \mnras, 393, 979

\bibitem[\protect\citeauthoryear{{Li}, {Colgate}, {Wendroff} \& {Liska}}{{Li}
  et~al.}{2001}]{LI01}
{Li} H.,  {Colgate} S.~A.,  {Wendroff} B.,    {Liska} R.,  2001, \apj, 551, 874

\bibitem[\protect\citeauthoryear{{Li}, {Finn}, {Lovelace} \& {Colgate}}{{Li}
  et~al.}{2000}]{LI00}
{Li} H.,  {Finn} J.~M.,  {Lovelace} R.~V.~E.,    {Colgate} S.~A.,  2000, \apj,
  533, 1023

\bibitem[\protect\citeauthoryear{{Lin}}{{Lin}}{2012}]{L12}
{Lin} M.-K.,  2012, \mnras, 426, 3211

\bibitem[\protect\citeauthoryear{{Lovelace} \& {Hohlfeld}}{{Lovelace} \&
  {Hohlfeld}}{1978}]{LOV78}
{Lovelace} R.~V.~E.,  {Hohlfeld} R.~G.,  1978, \apj, 221, 51

\bibitem[\protect\citeauthoryear{{Lovelace}, {Li}, {Colgate} \&
  {Nelson}}{{Lovelace} et~al.}{1999}]{LOV99}
{Lovelace} R.~V.~E.,  {Li} H.,  {Colgate} S.~A.,    {Nelson} A.~F.,  1999,
  \apj, 513, 805

\bibitem[\protect\citeauthoryear{{Lovelace}, {Turner} \& {Romanova}}{{Lovelace}
  et~al.}{2009}]{LTR09}
{Lovelace} R.~V.~E.,  {Turner} L.,    {Romanova} M.~M.,  2009, \apj, 701, 225

\bibitem[\protect\citeauthoryear{{Meheut}, {Casse}, {Varniere} \&
  {Tagger}}{{Meheut} et~al.}{2010}]{MEH10}
{Meheut} H.,  {Casse} F.,  {Varniere} P.,    {Tagger} M.,  2010, \aap, 516,
  A31+

\bibitem[\protect\citeauthoryear{{Meheut}, {Keppens}, {Casse} \&
  {Benz}}{{Meheut} et~al.}{2012}]{MKC12}
{Meheut} H.,  {Keppens} R.,  {Casse} F.,    {Benz} W.,  2012, \aap, 542, A9

\bibitem[\protect\citeauthoryear{{Meheut}, {Meliani}, {Varniere} \&
  {Benz}}{{Meheut} et~al.}{2012}]{MMV12}
{Meheut} H.,  {Meliani} Z.,  {Varniere} P.,    {Benz} W.,  2012, \aap, 545,
  A134

\bibitem[\protect\citeauthoryear{{Meheut}, {Yu} \& {Lai}}{{Meheut}
  et~al.}{2012}]{MYL12}
{Meheut} H.,  {Yu} C.,    {Lai} D.,  2012, \mnras, p.~2748

\bibitem[\protect\citeauthoryear{{O'Neil}}{{O'Neil}}{1965}]{ONE65}
{O'Neil} T.,  1965, Physics of Fluids, 8, 2255

\bibitem[\protect\citeauthoryear{{Sellwood} \& {Kahn}}{{Sellwood} \&
  {Kahn}}{1991}]{SEL91}
{Sellwood} J.~A.,  {Kahn} F.~D.,  1991, \mnras, 250, 278

\bibitem[\protect\citeauthoryear{{Tagger}}{{Tagger}}{2001}]{TAG01}
{Tagger} M.,  2001, \aap, 380, 750

\bibitem[\protect\citeauthoryear{{Tagger} \& {Melia}}{{Tagger} \&
  {Melia}}{2006}]{TAM06}
{Tagger} M.,  {Melia} F.,  2006, \apjl, 636, L33

\bibitem[\protect\citeauthoryear{{T{\'o}th}}{{T{\'o}th}}{1996}]{TOT96}
{T{\'o}th} G.,  1996, Astrophysical Letters Communications, 34, 245

\bibitem[\protect\citeauthoryear{{Tsang} \& {Lai}}{{Tsang} \&
  {Lai}}{2008}]{TSA08}
{Tsang} D.,  {Lai} D.,  2008, \mnras, 387, 446

\bibitem[\protect\citeauthoryear{{Umurhan}}{{Umurhan}}{2010}]{U10}
{Umurhan} O.~M.,  2010, \aap, 521, A25+

\bibitem[\protect\citeauthoryear{{Varni{\`e}re} \& {Tagger}}{{Varni{\`e}re} \&
  {Tagger}}{2006}]{VAR06}
{Varni{\`e}re} P.,  {Tagger} M.,  2006, \aap, 446, L13

\end{thebibliography}
\end{document}